\documentclass[letterpaper,12pt]{article}
\usepackage{graphicx}
\usepackage{setspace}

\newcommand{\ignore}[1]{}

\newcommand \EQ[2]
{
  \begin{equation} #1 \label{eq:#2}
  \end{equation}
}

\newcommand{\s}[1]{{\normalsize {\sf #1}}}
\newcommand{\n}[1]{{\small {\sf #1}}}

\begin{document}

\date{}

\title{Systematic Method for Path-Complete White Box Testing }


\author{
\begin{tabular}[t]{c@{\extracolsep{0.5in}}c}
Hanna Makaruk  & Robert Owczarek \\
\small{P-22, MS E548} & \small{NWO-TP, MS J594} \\
\small{Los Alamos National Laboratory} & \small{Los Alamos National Laboratory} \\
\small{Los Alamos, NM~~87545} & \small{Los Alamos, NM~~87545}\\
\small{hanna\_m@lanl.gov} & \small{rmo@lanl.gov}
\end{tabular}\\
Nikita A. Sakhanenko\\
\begin{tabular}[t]{c@{\extracolsep{0.5in}}c}
\small{P-22, MS E548} & \small{Computer Science Department} \\
\small{Los Alamos National Laboratory} & \small{University of New Mexico}\\
\small{Los Alamos, NM~~87545} & \small{Albuquerque, NM~~87131}\\
\small{nikita@lanl.gov} & \small{sanik@cs.unm.edu}
\end{tabular} }

\maketitle

\begin{abstract}
A systematic, language-independent method of finding a minimal set
of paths covering the code of a sequential program is proposed for
application in White Box testing. Execution of all paths from the
set ensures also statement coverage.
Execution fault marks problematic areas of the code.
The method starts from a UML activity diagram of a program.  The diagram
is transformed into a directed graph: graph's nodes substitute
decision and action points; graph's directed edges substitute action
arrows.

The number of independent paths equals easy-to-compute
cyclomatic complexity of the graph. Association of a vector to each
path creates a path vector space. Independence of the paths
is equivalent to linear independence of the vectors. It is sufficient
to test any base of the path space to complete the procedure.  An
effective algorithm for choosing the base paths is presented.

\vspace*{0.1in}\noindent
\end{abstract}

\doublespace

\section{Introduction}

For many years there have been theoretical studies of software, in
which the code structure is represented as a graph. In particular,
McCabe \cite{McCabe76} introduced cyclomatic complexity, known
from graph theory, as a measure of complexity of software. It is
also known \cite{McCabe76} that cyclomatic complexity describes
the number of independent paths through the code. Therefore,
cyclomatic complexity might be used in planning White Box software
testing. However, the practical side of this direction has not
been explored. There are some discouraging results from graph
theory telling that general problems of finding independent paths
are NP-complete (e.g. \cite{Erlebach} and references therein). On
the other hand, these complexity results are obtained for graphs
not necessary representing software. They are too general for the
software testing purposes. Software structure imposes important
constraints on the graph's structure and makes problem of finding
paths in such graph manageable. For example, every sequential
software has two characteristic nodes, the starting point and the
finishing point, and all the paths have to include these two
nodes. For convenience the starting point of a program is called
\s{begin} node, whereas the finishing point is called \s{end}
node, without any assumption about the programming language, in
which software is written.

In any graph corresponding to sequential code there is always a
path, which is the shortest (actually there may be more than one
such path, but one of them can be always chosen). In the paper
this path is called {\em backbone path}. The whole structure of
the code is built around this path. The graph structure of a
software code is relatively simple, usually with just a few
repeating constructs. As a result, it is possible to create an
algorithm, which allows for finding a set of independent paths
through arbitrary sequential code in a relatively easy way. This
paper presents new, more efficient approach to finding independent
paths and planning White Box testing.

There are many definitions of White Box software testing and its
completeness. Test completeness means that a predefined scope of
testing is exhausted. It is always a matter of choice which scope
of testing is needed, as no such thing as a perfect test,
disclosing every possible problem, exists. In this paper White Box
testing is called complete (path-complete) when software is
checked against major design flaws, the flaws are removed, and
then each line of the code is successfully executed at least once
during the test. In the literature (e.g. \cite{testing}) the
latter property is called {\em statement coverage}. Statement
coverage is usually considered as a minimum coverage requirement.
However, even some professional testers do not think it is
necessary to achieve this level of testing completeness.
Half-jokingly, Beizer \cite{testing} calls such approach
``criminal". The level of completeness achieved within the
proposed method is actually higher than statement coverage.
Checking against the major design flaws is in fact a prerequisite
in this method. Only software free of them can be properly tested.

The question how to test a known path has been already discussed
in \cite{GunterPeled99} resulting in a patented software
performing such tests \cite{patent}, thus, it is not covered in
this paper. From the paper by Gunter and Peled \cite{Gunter}, and
from private communication with one of them (Gunter) it seems that
the tool has some limitations, in particular with respect to data
types that might be entered. For the purpose of this paper it is
assumed that for any valid path through a program it is possible
to test it. Not only does the method presented in this paper
minimize a number of test paths needed to cover a program code,
but it also helps in identifying as short test paths as reasonably
achievable. Nevertheless, sometimes, when dealing with a
complicated real life software, one or more of the initially
identified test paths may be in fact very difficult or just
impractical to test. There is no need to spend excessive time on
dealing with such cases. The unusually difficult paths can be
substituted by others, according to the rules discussed in section
7.

The paper is outlined as follows.  The development of a method for
systematic White Box testing is discussed in the next sections: in
section 2 a control flow chart for a program is derived, and
checking against major design flaws is discussed; in section 3 a
graph description of the program is introduced, which enables
calculation of the number of independent paths in the program; in
section 4 a vector space of paths is defined and a simple method
for checking path independence is presented; in section 5 a
practical method for finding simplest independent paths is shown.
In section 6 an algorithm for test-planning is presented, and in
section 7 practical applications of a path vector space in the
test-planning are shown. Finally, in section 8 a summary of the
work and an overview of further research and possible applications
are given.

\section{Activity diagrams of a program and detection of major design flaws}

The first step of the proposed method is creation of a flow chart.
Any type of a flow chart can be used, if needed. In order to
standardize the approach and make it programming language
independent, a UML activity diagram is used in this paper. For
many programmers constructing an activity diagram is a routine
step in software design.  In case a program is written without
constructing an activity diagram at the design stage, it can be
reconstructed from the software code, as for each sequential
program a UML activity diagram can be constructed. Lines of the
code are assigned to decision or action points of its activity
diagram.

Before proceeding further, it is necessary to check the code
against major design flaws. Major design flaws include: parts of a
code that can never be executed, values that are calculated but
never used, redundant execution of the same piece of a code, or
any overcomplicated code, which can be easily refactored.

{\em Horrible} loops described in \cite{testing}, i.e. multiple
loops, which are neither nested nor consecutive, are clearly
identified by an activity diagram. Examples of {\em horrible}
loops include intersecting loops, cross-connected loops, jumps
into and out of loops, and hidden loops. Such constructions should
be removed from a program before further testing.

Procedures with multiple entrances and/or exits, which appear in
some old programs, for instance written in Fortran, may cause
another design problem. They are known also as {\em ill-formed
procedures} and their modification is generally advised
\cite{testing}. In case of multiple entrances an additional,
external entry point should be added. It is a decision point
directing control flow to the appropriate initial entry point.
Similarly, in case of multiple exits an external exit should be
added. All the initial exits would be then directed to it. Such a
modification of an {\em ill-formed procedure} is needed for
applying the next part of the method discussed here.

{\em Pachinko} code is an example of extreme control flow: a
program that modifies its own instructions. There is no known
method for testing it \cite{testing}, hence the only way is to
redesign such a program. Only software free of these flaws may be
properly tested.

{\em Spaghetti} code, in which control structure is tangled due to
multiple \s{go-to} jumps and other branching constructs, is not a
mark of good contemporary design. Nevertheless, if it is free of
{\em horrible} loops, it still might be tested by the method
proposed further. On the other hand, redesigning {\em spaghetti}
code is strongly suggested by software designers that can be
achieved easily using activity diagram.

\section{Graph representation of a program, independent paths}

After creating an activity diagram for a code and removing any
design flaws found, the next step is to create a directed graph
representing the control flow in the code. The directed graph is
built from the UML activity diagram in the following way: graph
nodes are substituted for all the decision and action points,
directed edges of the graph are substituted for the action arrows.
Such a graph contains a single input node (\s{begin}) and a single
output node (\s{end}). A virtual edge from \s{begin} node of the
graph to \s{end} node is used to close the directed graph. This
way the graph becomes strongly connected.

\noindent
{\bf Definition. }A {\em strongly connected graph} is a graph,
whose any two nodes are connected by a {\em path}, which is a
sequence of consecutive directed edges.

From the graph topology point of view, independent paths closed by
the same virtual edge can be understood as cycles constituting the
first homology space for the graph supplemented by the virtual
edge. Hence cyclomatic complexity, which is of basic importance
for this work, is equal to the first Betti number for the graph
(e.g. \cite{Duke}).

The number of independent paths in the whole program as well as
the number of independent paths inside any module or inside a
complex loop can be established just by counting the number of
edges and nodes of the corresponding graph. It is one of the
measures of software complexity. It has an advantage over
language-dependent or even programming-style-dependent code
measures like the number of lines, because it is related to
logical structure of a program (decisions, loops).

For a strongly connected graph the number of independent paths is equal
to the cyclomatic complexity number of the graph. The complexity number,
$V(G)$, of a graph $G$ is defined as
\EQ{
V(G)= e-v+1,}{cycl} where $e$ is the number of edges in the graph,
and $v$ is the number of its nodes. A virtual edge should be added
to make the obtained graph strongly connected. Therefore, the
number of edges is increased by one, and (\ref{eq:cycl}) becomes
$V(G)=(e+1)-v+1$.

Transition from an activity diagram to a directed graph preserves
paths. It is desirable to use independent paths in software
testing for computational efficiency. In the next section of the
paper, a vector representation of the paths is introduced. Path
independence can be checked easily using this vector description.

\noindent
{\bf Definition. }A path $p$ is {\em independent} from a set $S$
of other paths if for any path $q \in S$, $p \ne q$, there exists
a part of a code executed by the path $p$ such that it is never
executed by the path $q$.

Loops and jumps are special constructs in software, that is why
their representation should be discussed in more detail. In case
of consecutive loops, each of them is treated separately; nested
loops are treated as subgraphs. At this stage of the method the
code is expected to be already free of all other multi-loop
constructions, like intertwining loops or hidden loops
\cite{testing}.

In the loops of the most common type a condition is evaluated
before execution of the loop. A typical example is \s{for}
construct. According to the convention introduced earlier, it is
sufficient to traverse each edge of a graph once to reach
completeness of the testing. In other words it is sufficient to
traverse a loop once, if it does not have internal decision points
or internal loops. A subgraph of a loop with internal structure
can be considered separately from the main graph such that a
condition-checking vertex becomes both \s{begin} and \s{end}
nodes. The number of independent paths needed to test such a loop
is determined by structure of this subgraph.

There is also another type of loops, in which the condition is
evaluated at the bottom of the loop, for instance \s{do-while}
construct. A body of such a loop is executed once before the
condition is ever checked. After that the loop execution is done
similarly to \s{for} loop. In other words, a \s{do-while} loop can
be represented via \s{while} loop (which is equivalent to \s{for}
loop) as follows:

$$
\begin{array}{rrr}
\begin{minipage}{1.5in}
\s{do \{}\\
\hspace*{0.3in} \s{do-while body;}\\
\s{\} while(} cond \s{);}
\end{minipage} &
\Longrightarrow &
\begin{minipage}{1.5in}
\s{do-while body;}\\
\s{while(} cond \s{);}\\
\hspace*{0.3in} \s{do-while body;}
\end{minipage}
\end{array}
$$
\noindent
This formalism shows that only one type of loops is needed for a
graph representation.

A construct called \s{go-to} jump is not recommended in modern
programming. This construct distorts a coding block and hence
eliminates any visual cues about the flow of control. Even though
it can be avoided, some programmers still use \s{go-to}.
Therefore, in order for the method to be complete it should be
able to handle this construct as well. In case of a forward
\s{go-to} jump, it can be represented as an additional edge
between \s{go-to} and its label. Otherwise, if the jump is
backward, it can be represented similarly to \s{do-while} loop
discussed above.

There is yet another issue to consider: is it sufficient to
traverse a loop only once during testing? For finding paths, only
one execution of a simple loop (a loop without internal loops or
decision points) is assumed. The problem of a number of executions
of a loop is not handled by presented method, since finding the
paths is the focus of this paper, not their testing. Once the loop
path is established, it can be tested as needed. Usually it is
much easier to do so after the path structure for the whole
program is identified. \begin{figure}[h] \centering
\fbox{\includegraphics[scale=0.48]{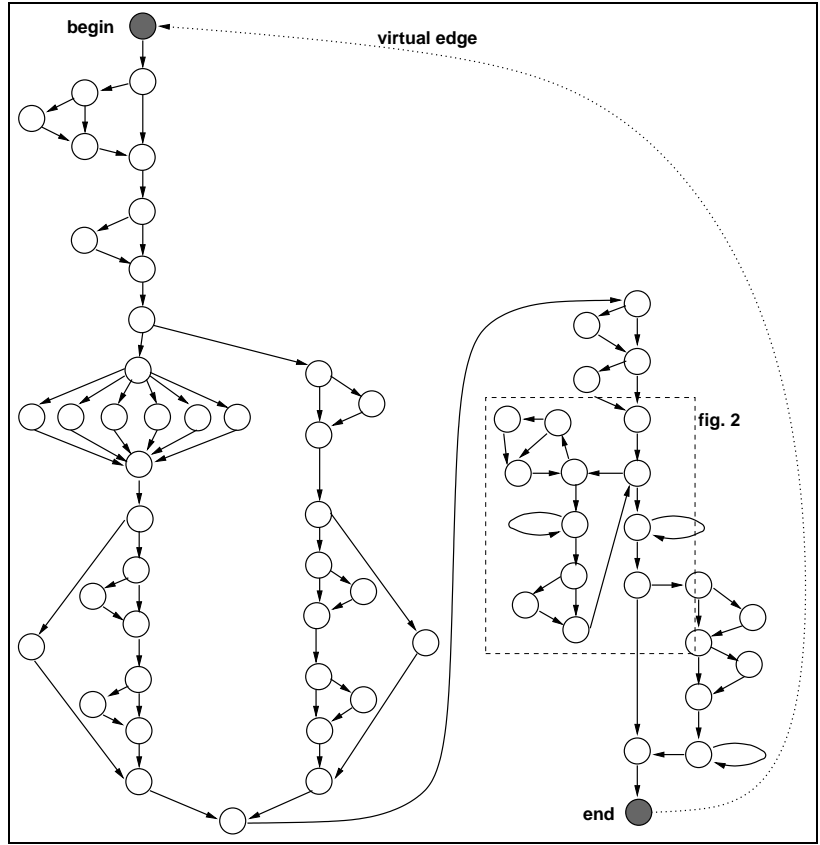}}
\caption{{\footnotesize Directed graph of ProtoReX: Prototype
Software for Objects Reconstruction from X-ray Images (ProtoReX).
Developed by T. Asaki, C. Powell, H. Makaruk at LANL, 2004,
unpublished; implemented in MatLab.}} \label{fig:big}
\end{figure}
Figure \ref{fig:big} shows an example of a graph constructed from
the software for object reconstruction from X-ray images.

\section{Vector representation of paths}

In order to check path independence a vector is assigned to each
possible path. The number of elements of each vector equals the
number of edges in the graph. Note that edges have to be ordered,
however the order can be arbitrary. Each element of the vector
corresponding to a particular edge indicates whether the edge
belongs to the path or not.  The element is equal to 1 if the edge
belongs to the corresponding path or 0 otherwise.

Since elements of the vector can only be 0 or 1, the vector space
of paths is defined over {\sf Z\hspace{-1.6mm}Z}$_2$ (the field of
integers modulo 2). Linearly independent vectors represent
independent paths, which enables checking paths' independence by
showing linear independence of the vectors representing them. Even
though the vectors' length is large, their elements are only zeros
or ones. Moreover all calculations are done modulo 2 that is why
the computation is still simple.

The dimension of a path vector space is equal to the number of
independent paths in a graph, i.e. to cyclomatic complexity $C =
e-v+2$ of the graph. Therefore, any set of C vectors representing
independent paths forms a base in this vector space.

It is sufficient to restrict testing only to the base paths in
order to guarantee that all the nodes are visited, i.e. to achieve
statement coverage \cite{testing}. In fact with this method a path
coverage is achieved, which is more than just statement coverage.
A choice of the base paths is obviously not unique. For the
practical reasons it is very important to choose the base in a
convenient way. A method for choosing the base is discussed below.
Checking for linear independence of appropriate vectors is
sufficient for showing independence of corresponding paths.

This concludes the presentation of a theoretical basis of a
general procedure. The following section of the paper introduces a
method for effective choice of the base paths. This method is
powerful enough to handle relatively complex software. It is
oriented towards choosing a set of the simplest independent paths
among different possibilities.

\section{Method for effective selection of the base paths}

\subsection{Backbone path}

Each path in a graph constructed accordingly to the rules
described above connects \s{begin} and \s{end} nodes. Recall that
a path that contains the minimum number of edges is called a
backbone path. It is possible and, in fact, not uncommon that more
than one path with the minimum number of nodes exists; any one of
them can be chosen to be a backbone path. The only fact important
for further consideration is that for any sequential software it
is always possible to choose a backbone path. Choosing the
simplest possible path is not a strict requirement, however it
simplifies all the next steps of the method.

\subsection{Loops}
\label{subsec:loops}

\begin{figure}[h]
\centering \fbox{\includegraphics[scale=0.65]{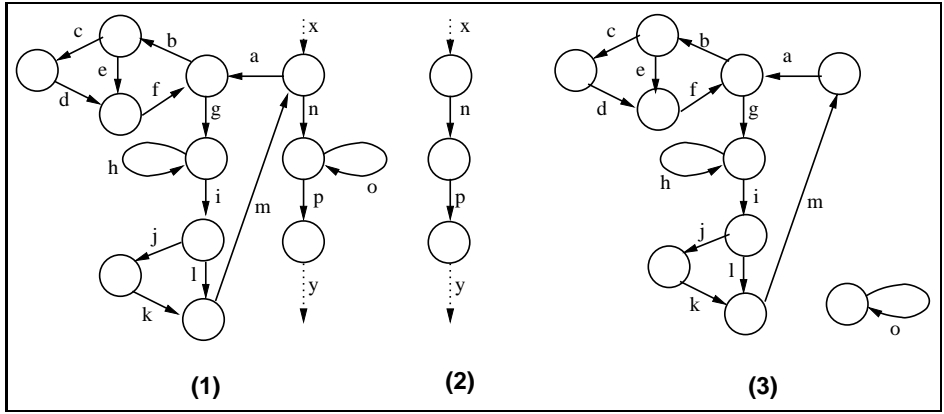}}
\caption{{\footnotesize (1) A part of ProtoReX graph; (2) its
loopless part (backbone); (3) its loop-graphs. }} \label{fig:part}
\end{figure}

Loops are special components of a graph. Any graph can be
represented as a merge of its loopless part and loop-graphs. A
loopless part of a graph is obtained by removing all its loops,
whereas a loop-graph is a subgraph, which represents a loop of
arbitrary complexity (see figure \ref{fig:part}).

A loop can originate inside another loop or block (loops {\em (h)}
and {\em (bef)} of figure \ref{fig:part}) or on a backbone path
(loop {\em (o)} of figure \ref{fig:part}). In this construction
nested loops are represented by the outermost loop. If a loop
originates on a backbone path, its {\em extended loop-graph}
consists of the loop added to the backbone (path {\em xn(o)py} for
loop {\em (o)}; parentheses indicate looping inside of a path).
Otherwise, if a loop originates outside of a backbone path (inside
of a block) an extended loop-graph consists of the loop with the
shortest path between \s{begin} and \s{end} of a block added to a
part of the backbone path outside this block (path {\em
xa(bef)gilmnpy} for loop {\em (bef)}).

After this reduction, the vector space of all possible paths can
be represented as a direct sum of the space for the loopless part
of the graph and each of the spaces for extended loop-graphs
described above. Stated another way, if $S_0$ is a vector space of
the loopless part, $S_i$ is a vector space of paths of the
$i^{th}$ extended loop-graph, where $i = 1, \ldots ,n$ and $n$ is
the number of loops, then vector space $S$ of paths of the graph
is
$$
S = \bigoplus_{i=0}^n S_i.
$$
A simple loop, which does not have nested loops or conditionals,
can be tested by executing a single path passing through the loop.
On the other hand, a complex loop can be considered separately,
since it is represented by a strongly connected subgraph with
single \s{begin} and \s{end} nodes (it is strongly connected since
a virtual edge from {\sf end} to {\sf begin} is assumed). This
method can be applied recursively on this subgraph to obtain its
independent paths.

\subsection{Modularity of a program}

In this subsection loopless part of a graph is considered. It is
common that graphs of sequential programs, especially long ones,
contain cut-vertices.

\noindent
{\bf Definition. }A {\em cut-vertex} of a graph, also known as an
{\em articulation point}, is a node whose removal disconnects the
graph.

This definition implies that every path in a software graph passes through
every cut-vertex of the graph.

\noindent
{\bf Definition. }A part of a graph between two cut-vertices is called a {\em module}
in software development and a {\em block} in graph theory.

A block can also be closed by a virtual edge,
hence the number of independent paths in a block can be calculated the same
way as for the whole graph.

Each block can be considered as a subgraph with its own cyclomatic
complexity number and its own set of paths. A section of a
backbone path belonging to the block should be identified as one
of the base paths in this block.  When a base of a block path
space is established, all the paths are complemented by segments
of the backbone path outside of the block (see subsection
\ref{subsec:loops} for details). \ignore{In order to establish a
base of the space of paths for the block, one can recursively
apply this procedure to its sub-blocks. It is easy to find a set
of independent paths of the smallest sub-block (sub-blocks of
which are trivial), which constitutes the base case of this
recursion.}

A complex block can be divided into two or more parallel parts. A
path space of a block is a sum of path spaces of all the parts. In
case any part has a cut-vertex(ices) it can be divided into its
own sub-blocks. Division into smaller and smaller sub-blocks can
be repeated until the smallest units are reached, sub-blocks of
which are trivial.

Some remarks about how to build a base of the path space for the
whole graph from bases of path spaces for its blocks and
loop-graphs should be made. Every block shares a backbone path
with other blocks, hence one of the base paths of each module is a
segment of the backbone path. Any two base paths with non-backbone
segments included in different blocks are independent.

Merging all bases of block paths and loop paths can serve as a
base of the graph. A merge operation here is a set union,
therefore, a backbone path appearing in the base of each block is
included only once.

Not only does this method make the process of finding the base paths
relatively simple, but it also constructs paths which are generally shorter
than their possible alternatives. In addition there are several other
advantages of the method.

One of them is better localization of the problem in the code than
the one provided by traditional testing methods. In case one of
the base paths fails the test, the failure can be localized to the
part of the code corresponding to this path, and frequently even
to a part of this path in a particular module.

It is also important that a recursive approach allows for applying the method
effectively to a sequential program of practically any complexity.

\section{Path-choosing algorithm}

In this section an effective algorithm for path finding is proposed. It can
be applied to the whole program as well as to its submodules. The method, called
\n{TEST}, is presented in a pseudo-code.

\singlespace
\noindent
\n{TEST(C):}\\
\n{INPUT:} software code \n{C}.\\
\n{OUTPUT:} set \n{P} of all independent paths.\\
\n{BEGIN}\\
\indent\indent \n{1. A}\s{=get\_activity\_diagram(}\n{C}\s{);}\\
\indent\indent \n{2. A}\s{=purify(}\n{A}\s{);}\\
\indent\indent \n{3. G}\s{=get\_graph(}\n{A}\s{);}\\
\indent\indent \n{4. }\s{n=get\_number\_indep\_paths(}\n{G}\s{);}\\
\indent\indent \n{5. P}\s{=compute\_paths(}\n{G}\s{);}\\
\n{END}

\doublespace
\noindent
In the first step of the algorithm a UML activity diagram of the
software denoted as \n{C} is prepared. Function \s{purify}, which
is called afterwards, eliminates design flaws, if any, and
optimizes the software design, if applicable. After that the
activity diagram is translated into a directed graph by calling
function \s{get\_graph}. Finally, the number of independent paths
(read cyclomatic complexity) through the connected graph obtained
in the previous step is calculated and function \s{compute\_paths}
is called. The latter function is the actual algorithm for finding
independent paths of a strongly connected graph. Its pseudo-code
is given below.

\singlespace
\noindent
\s{compute\_paths(}\n{C}\s{):}\\
\n{INPUT:} directed, strongly connected graph \n{G}.\\
\n{OUTPUT:} set \n{P} of all independent paths of graph \n{G}.\\
\n{BEGIN}\\
\indent\indent \n{1. P}\s{=}\n{$\emptyset$}\s{;}\\
\indent\indent \n{2. }\s{loop}\\
\indent\indent\indent\indent \n{2.1. M$_i$}\s{=get\_block(}\n{G}\s{);}\\
\indent\indent\indent\indent \n{2.2. }\s{if(}\n{M$_i$==G}\s{)}\\
\indent\indent\indent\indent\indent\indent \n{2.2.1. P$_i$}\s{=count(}\n{M$_i$}\s{);}\\
\indent\indent\indent\indent \n{2.3. }\s{else}\\
\indent\indent\indent\indent\indent\indent \n{2.3.1. P$_i$}\s{=compute\_paths(}\n{M$_i$}\s{);}\\
\indent\indent\indent\indent \n{2.4. P}\s{=combine\_paths(}\n{P,P$_i$}\s{);}\\
\indent\indent \n{3. }\s{loop until(}no sub-blocks left\s{);}\\
\n{END}

\doublespace
\noindent
Function \s{compute\_paths} is a recursion: it recursively
computes independent paths of each sub-block of the given graph
and then combines these sets together. Function
\s{combine\_paths}, which performs such combining, is given below.
Note also that the base case of the recursion is computed by
\s{count}, which is a simple enumeration of independent paths in
the trivial sub-block.

\singlespace
\noindent
\s{combine\_paths(}\n{P$_1$,P$_2$}\s{):}\\
\n{INPUT:} two sets of independent paths: \n{P$_1$} and \n{P$_2$}.\\
\n{OUTPUT:} set \n{P} of independent paths created from \n{P$_1$} and \n{P$_2$}.\\
\n{BEGIN}\\
\indent\indent \n{1. }\s{b$_1$=get\_backbone(}\n{P$_1$}\s{);}\\
\indent\indent \n{2. }\s{b$_2$=get\_backbone(}\n{P$_2$}\s{);}\\
\indent\indent \n{3. }\s{for(}each path \s{a$_1$}$\in$\n{P$_1$}: \s{a$_1$}$\ne$\s{b$_1$)}\\
\indent\indent\indent\indent \n{3.1. P}\s{=add(}\n{P}\s{,a$_1$}$\cdotp$\s{b$_2$);}\\
\indent\indent \n{4. }\s{for(}each path \s{a$_2$}$\in$\n{P$_2$}: \s{a$_2$}$\ne$\s{b$_2$)}\\
\indent\indent\indent\indent \n{4.1. P}\s{=add(}\n{P}\s{,b$_1$}$\cdotp$\s{a$_2$);}\\
\n{END}

\doublespace
\noindent
The idea of function \s{combine\_paths} is to append each path
from \n{P$_1$} with a backbone path of \n{P$_2$} creating a subset
of paths. Similarly, each path from \n{P$_2$} is appended to a
backbone path of \n{P$_1$} creating another subset of paths, which
is then merged with the first one. Operation "$\cdotp$" in the
program above is an append of two paths. Note that the paths order
during the append is important. Function \s{add} in the program is
an inclusion of a given path into a path set.
\begin{figure}[h]
\centering \fbox{\includegraphics[scale=0.65]{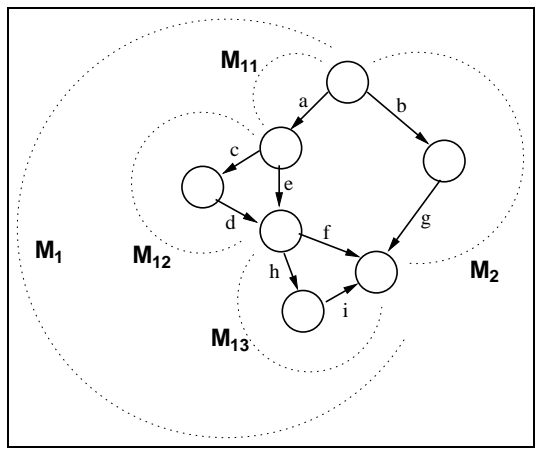}}
\caption{{\footnotesize An example of a graph with block
decomposition. }} \label{fig:example}
\end{figure}
As an example function \s{compute\_paths} is applied to a graph in figure \ref{fig:example}.
The graph is divided into two blocks: \n{M$_1$} and \n{M$_2$}, and the function is
recursively called on each block. In case of the block \n{M$_1$}, it is further divided into
\n{M$_{11}$}, \n{M$_{12}$}, and \n{M$_{13}$}. Block \n{M$_{11}$} cannot be divided anymore,
hence \n{P$_{11}$}\s{=count(}\n{M$_{11}$}\s{)=\{a\}}. A recursive call of \s{compute\_paths}
on \n{M$_{12}$} returns \n{P$_{12}$}\s{=\{e,cd\}} which needs to be combined with
\n{P}\s{=}\n{P$_{11}$}. Combined \n{P} trivially equals to \s{\{ae,acd\}}. Similarly,
\n{P$_{13}$}\s{=compute\_paths(}\n{M$_{13}$}\s{)=\{f,hi\}}. While combining \n{P$_{13}$}
with current \n{P}, the backbone of each set is identified: \s{f} and \s{ae} correspondingly.
The combination of two sets is \n{P}\s{=\{aef,aehi,acdf\}}. Consequently, the final
set of independent paths for the graph from figure \ref{fig:example} is
\n{P}\s{=\{aef,aehi,acdf,bg\}}.

\ignore{Remark: Some experience with the method allows a tester to easily find
all independent paths through a relatively complex module on the base of intuition, without need
for dividing it into sub-modules.}

This algorithm finds the minimal number of paths and constructs
paths needed to be tested for a complete (in the sense of
path-completeness) White Box testing. \ignore{The paths
constructed by the algorithm are the shortest possible independent
paths, if properly chosen inside each single module.} Since each
path diverts from the backbone path inside of only one module
loop, it might be relatively easy to localize the problem to a
particular module/loop when a path fails a test. Tracing where the
path diverts from other paths inside the same module can localize
the problem area even further.

\section{Applications of vector space in path finding}

Vectors can be assigned to each path in the way described before.
Checking a number of the paths established using the path-finding
algorithm as well as linear independence of their vectors
independently verifies that test paths are established correctly.
In case of any possible mistake checking the number of independent
paths contributed by each of the blocks and loops quickly
localizes the problem inside one of them.

A vector space of all paths through the graph $G$, $V(G)$ can be
represented as a simple sum of a vector space of paths for
loopless part of the graph $V(G_0)$ and all of the extended loop
graphs $V(G_l)$, where $n_l$ is the number of extended loop
graphs:
\EQ{
V(G)= V(G_0)\oplus\sum\limits_{l=1}^{n_l} V(G_l).}{vecspace} All
blocks, on the other hand, share a backbone path, which is one of
the base paths for each of them. This is why each block
contributes one path less than the dimension of its paths vector
space, and the backbone path is counted once:
\EQ{
C_g=1+\sum_{b=1}^{n_b}(C_b-1)+\sum_{l=1}^{n_l}C_l.}{count}
where $C_g$, $C_b$, and $C_l$ represent complexity numbers of the whole graph (equals $e-v+2$),
a loopless block, and an extended loop graph correspondingly.

An experienced tester dealing with a relatively simple graph (a
module, a loop) may not need to apply the algorithm at all, but
use the intuition to guess an appropriate number of paths and
check linear independence of their vectors. In cases when it is
applicable, such an approach simplifies the method even more.

It may happen that some of the initially identified base paths are
difficult or practically impossible to test. In this situation a
new base is established by supplementing the to-be-tested base
paths with a set of paths, which are different from rejected ones,
but independent from the other base paths. Substituting two or
more base paths by more convenient for testing paths may be a time
saving strategy, while preserving test-completeness.

\section{Conclusion}

In this paper, using graph theory, an effective algorithm for
finding sets of independent paths in sequential software is found.
In addition, an efficient method for independence checking based
on linear algebra is given. The method finds a base in the set of
independent paths, and shows that the resulting set of paths
constitutes a base. Testing any base paths completes White Box
testing. Therefore, proposed algorithm allows for planning White
Box software testing in a systematic way, and for performing such
testing in a complete way in the sense of path completeness.

There are two possible directions for the future work. The first
direction is to implement the White Box testing method described
in the paper. The second is to generalize the method to handle
parallel programs. A UML activity diagram can be constructed for
parallel software, which in turn can be transformed into a
directed graph similarly to the sequential case. The main
challenge is that in general parallel software has to be tested
not only for execution of every possible path but also for
synchronization between the executions of different parts of the
code.

\section{Acknowledgements}
The research was connected to work on an X-ray image recognition
project. The authors are grateful to members of the Data Driven
Modelling and Analysis team for their help, particularly to Tom
Asaki for his hospitality, discussions and enthusiasm. The help of
Collin Powell for providing the ProtoReX MatLab code, on which
initial ideas were tested, is acknowledged.

The authors are thankful to David Aubrey for extensive discussions
and many practical suggestions about software testing.

Many thanks to Josef Przytycki for discussions and checking the
graph-theoretical basis of the method.

\end{document}